\documentclass[aps,prl,superscriptaddress,twocolumn]{revtex4}

\usepackage{graphicx} 
\usepackage{amsmath}
 
\newcommand{\beq}{\begin{equation}}
\newcommand{\enq}{\end{equation}}
\newcommand{\la}{\langle}
\newcommand{\ra}{\rangle}

\begin{document}

\title{Quasi two-dimensional superfluid Fermi gases}
\author{J.-P. Martikainen}
\affiliation{Nanoscience Center, Department of Physics,
PO Box 35, 40014 University of Jyv\"{a}skyl\"{a}, Finland}
\author{P. T\"{o}rm\"{a}}
\affiliation{Nanoscience Center, Department of Physics,
PO Box 35, 40014 University of Jyv\"{a}skyl\"{a}, Finland}
\affiliation{
Institute for Quantum Optics and Quantum Information of the Austrian Academy of Sciences, A-6020 Innsbruck, Austria}
\date{\today}

\begin{abstract}
We study a quasi two-dimensional superfluid Fermi gas where
the confinement in the third direction is due to
a strong harmonic trapping. We investigate the behavior
of such a system when the chemical potential is
varied and find strong modifications of the superfluid properties
due to the discrete harmonic oscillator states.
We show that such quasi two-dimensional behavior
can be created and observed with current experimental capabilities.
\end{abstract}
\maketitle 

Finite size effects often lead to pronounced 
quantum mechanical changes in the system behavior.
Among other things, these effects are crucial in the physics of quantum dots~\cite{Reimann2002a},
in studying how spontaneous emission is modified 
in cavities~\cite{Hulet1985a,Heinzen1987a,Gabrielse1985a},
and in understanding the properties of thin
superconducting films~\cite{Guo2004a}.
In addition, dimensional cross-over effects have been actively studied
experimentally using ultracold bosons in combined magnetic and optical potentials~\cite{Goerlitz2001a}
as well as in optical 
lattices~\cite{Greiner2001a,Greiner2001b,Paredes2004a}.

The possibility of fermionic superfluidity in ultracold atomic
gases~\cite{Stoof1996b} motivated a substantial experimental effort 
into controlling and manipulating cold fermionic atoms.
Very recently, these efforts were rewarded with a spectacular success.
A series of experiments demonstrated Bose-Einstein condensation of molecules
composed of two fermions~\cite{Jochim2003b,Greiner2003a,Zwierlein2003a,Bourdel2004a}.
Experiments~\cite{Regal2004a,Zwierlein2004a,Kinast2004a,Bartenstein2004b}
employing the Feshbach resonance to vary the atomic interaction strength
observed strong indications of fermion pairing. This was soon followed by the direct observation of the energy
gap~\cite{Chin2004a}. This set of experiments have established 
fermionic pairing in atomic gases as an experimental fact, and provided a strong case for 
superfluidity, proven recently by the observation of vortices~\cite{Ketterle2005}.

The dramatic progress with ultracold fermions, combined with the fact
that there is no fundamental problem in changing the 
dimensionality of the system by using for example optical lattices~\cite{Kohl2005a},  
raises important questions about the role of
dimensionality and finite size effects in cold fermionic gases.
The purpose of this Letter is to elucidate how 
quantum size effects are manifested 
in a dilute quasi-two-dimensional Fermi gas.
Such system can be created by confining atoms tightly along one direction by
a harmonic trapping potential. Note that we do not assume a purely
Gaussian density profile in the $z$-direction, an assumption
often made in studies of quasi two-dimensional quantum
degenerate gases. Our extension beyond the ground state Gaussian profile reveals discrete features
in observable quantities, and is expected to be useful in
studying how a true long range order at non-zero temperature is established as one
crosses over from a purely two-dimensional into a three-dimensional system.

We assume a cloud of fermionic atoms confined by 
a potential $V(z)=m\omega_z^2 z^2/2$ and
an atomic density $n(z)$ which only depends on the $z$-coordinate.
The cloud consists of equal amount of fermions on two different 
internal states denoted by $\uparrow$ and $\downarrow$. 
Let us first, for orientation, consider the behavior of
a non-interacting system at zero temperature: 
The first atom inserted into the system will fill the
lowest energy state which corresponds to a stationary atom 
in the $xy$-plane and whose axial
wave function is the lowest harmonic oscillator state 
$\phi_0(z)$ of the axial potential. This atom then has
the energy $\hbar\omega_z/2$. As we add more atoms,
they fill the continuum of plane-wave states
in the $xy$-plane with an axial wave function $\phi_0(z)$.
However, once the two-dimensional density of atoms
is such that the next available plane-wave state has an energy
$\hbar\omega_z$, the density of states changes abruptly and 
an extra atom has two choices. Either,
it can occupy this high energy plane-wave state or
it can occupy the zero plane-wave momentum state in the first
excited harmonic oscillator state $\phi_1(z)$. Similar doubling
of choices occurs also with respect to higher harmonic oscillator
states, when the continuum Fermi-energy matches the oscillator
level energy.
More quantitatively, the two-dimensional density (of both components
combined) $n_{2D}=\int dz\, n(z)$ is related
to the Fermi-momentum through $k_F^2=2\pi n_{2D}$. If the
Fermi-energy $E_F=\hbar^2k_F^2/2m$ is equated with the level spacing
$\hbar\omega_z$ we find a maximum density before atoms
start to occupy the first excited state of the harmonic oscillator,
$n_c=1/(\pi l_z^2)$. (If not indicated otherwise, we use $\hbar\omega_z$ as a 
unit of energy and $l_z=\sqrt{\hbar/m\omega_z}$ as a unit of length.)
As the density exceeds this threshold, extra atoms have
available a new state with the same total energy as the plane-wave
states, but with a different axial state. This indicates that
the growth rate of $n_{2D}$ with $E_F$ is suddenly doubled from
its earlier value of $\partial n_{2D}/\partial E_F=m/(\pi\hbar^2)$.
Therefore, $\partial n_{2D}/\partial E_F$ for
ideal fermions, at zero temperature, is a staircase with steps
at $E_F=\hbar\omega_z\left(n+1/2\right)$ with the height $m/(\pi\hbar^2)$.

Let us now consider an attractive binary contact interaction with a
coupling strength $g$ between
atoms in different internal states and allow for a non-zero temperature. 
Interaction between atoms is assumed to be sufficiently weak so that
the Bardeen-Cooper-Schrieffer (BCS) theory provides a reliable framework to study this many-body system
also at finite temperature. 
If the Hamiltonian is expanded around the order parameter
$\Delta(z)=g\la \hat{\psi}_\downarrow({\bf r})\hat{\psi}_\uparrow({\bf r})\ra$, one finds the usual quadratic
mean field Hamiltonian 
\begin{eqnarray}
H&=&\int d{\bf r}\sum_{\sigma} \hat{\psi}_\sigma^{\dagger}({\bf r})
\left(-\frac{\hbar^2}{2m}\nabla^2-\mu+V(z)\right)\hat{\psi}_\sigma({\bf r})
\nonumber\\
&+&\Delta({\bf r})\hat{\psi}_\uparrow^{\dagger}({\bf r})\hat{\psi}_\downarrow^{\dagger}({\bf r})
+\Delta^*({\bf r})\hat{\psi}_\downarrow({\bf r})\hat{\psi}_\uparrow({\bf r}) .
\end{eqnarray}
This Hamiltonian is diagonalized using a
Bogoliubov transformation
$\hat{\psi}_\uparrow({\bf r})=\sum_\zeta u_\zeta({\bf r})\hat{b}_{\zeta,\uparrow}
+v_\zeta^*({\bf r})\hat{b}_{\zeta,\downarrow}^\dagger$ and 
$\hat{\psi}_\downarrow^\dagger({\bf r})=\sum_\zeta -v_\zeta({\bf r})\hat{b}_{\zeta,\uparrow}
+u_\zeta^*({\bf r})\hat{b}_{\zeta,\downarrow}^\dagger$,
where the quasiparticle amplitudes $u_\zeta({\bf r})$ and $v_\zeta({\bf r})$ are solutions to the 
Bogoliubov-de-Gennes (BdG) equations
\beq 
\left(\begin{array}{cc} H_0 & \Delta({\bf r})\\
\Delta^*({\bf r}) & -H_0\end{array}\right)
\left(\begin{array}{c} u_\zeta({\bf r}) \\ v_\zeta({\bf r})\end{array}\right)=
E_\zeta\left(\begin{array}{c} u_\zeta({\bf r}) \\ v_\zeta({\bf r})\end{array}\right),\nonumber\\
\enq
where $H_0=-\hbar^2\nabla^2/2m-\mu+V(z)$.
Furthermore, the amplitudes are normalized 
$\int |u_\zeta({\bf  r})|^2+|v_\zeta({\bf r})|^2=1$.
Self-consistency then imposes the well known
gap equation
\beq
\label{eq:gapeq}
\Delta(z)=-g\sum_\zeta u_\zeta({\bf r})v_\zeta^*({\bf r})\left[
1-2n_F(E_\zeta)\right],
\enq
where $n_F(E)=1/\left(\exp(\beta E)+1\right)$ is the Fermi distribution. 
Finally, the chemical potential is related to the atom density through
the number equation
$n({\bf r})=\sum_\zeta |u_\zeta({\bf  r})|^2n_F(E_\zeta)+
|v_\zeta({\bf r})|^2\left(1-n_F(E_\zeta)\right).
$

In the quasi-two-dimensional system considered here, it is natural to expand the
Bogoliubov quasiparticle amplitudes in terms of the 
harmonic oscillator states $\phi_n(z)$ and radial plane waves
$\sim\exp\left(i{\bf k}\cdot{\bf r}_\perp\right)$. This amounts to
\beq
u_\zeta({\bf r})=\sum_n\sum_{\bf k} \frac{1}{\sqrt{A}}\phi_n(z)
e^{i{\bf k}\cdot{\bf r}_\perp} u_{n,k}^\zeta 
\enq
and
the same expression for $v_\zeta({\bf r})$, 
where $A$ is the quantization area in the $xy$-plane.
We include only the three lowest lying harmonic oscillator states
and the solution to the BdG equations for 
the amplitudes $u_{n,k}^\zeta$ and $v_{n,k}^\zeta$
amounts to a diagonalization of the matrix
\beq
M=\left(\begin{array}{cccccc}
\xi_{0,0,k} & \Delta_0 &0&0 &0& \Delta_{02}\\
\Delta_0& -\xi_{0,0,k} & 0& 0& \Delta_{02}&0\\
0& 0& \xi_{0,1,k}& \Delta_1 &0&0\\
0 &0& \Delta_1&-\xi_{0,1,k}&0&0\\
0& \Delta_{02} &0&0& \xi_{0,2,k}&\Delta_2\\
\Delta_{02} & 0& 0&0& \Delta_2 & -\xi_{0,2,k}
\end{array}\right)\nonumber,
\enq
where $\xi_{0,n,k}=\hbar\omega_z(n+1/2)+\hbar^2k^2/2m-\mu$, $\Delta_n=\int dz
\Delta(z)|\phi_n(z)|^2$, and 
$\Delta_{02}=\int dz \Delta(z)\phi_2^*(z)\phi_0(z)$. 
This matrix is almost block-diagonal with respect
to different harmonic oscillator states. However, 
since $\phi_0(z)$ and $\phi_2(z)$ are both symmetric, $\Delta_{02}$
is non-zero and the  simple block diagonality is broken. This coupling
between $n=0$ and  $n=2$ channels is evident as an 
avoided crossing between two of the three (positive) dispersion branches.

For a weakly interacting system the restriction to just three harmonic oscillator states
is expected to be sufficient, if the chemical potential 
is below the energy $7\hbar\omega_z/2$ of the third excited state and 
if the temperature is small compared to $\hbar\omega_z$. 
In our examples both these conditions are well satisfied.
The coupling between $n=0$ and $n=2$ states influences, depending on $\mu$, $\Delta(z=0)$ by about $10\%$. 
As a validity test, we included the $n=3$ channel in the numerics and found a quantitative change of a few percent, 
while the qualitative behavior was unchanged.

The eigenvectors
${\bf w}_{\zeta,k}=\left(u_{0,k}^\zeta ,v_{0,k}^\zeta ,u_{1,k}^\zeta
  ,v_{1,k}^\zeta ,u_{2,k}^\zeta ,v_{2,k}^\zeta \right)$ 
with eigenvalues $E_{\zeta,k}$
can then be inserted into Eq.~(\ref{eq:gapeq})
to find a self consistent solution for $\Delta(z)$.
Since the Bogoliubov quasiparticle amplitudes are polynomials multiplied by
the same exponential (in trap units) $\exp(-z^2/2)$,
it is convenient to expand the symmetric order parameter as
$\Delta(z)=\sum_{n=0}^2 \alpha_{2n} z^{2n}\exp(-z^2)$. 
When the above expansion is used in Eq.~(\ref{eq:gapeq}) in conjunction
with eigenvectors ${\bf w}_{\zeta,k}$, we find, by comparing terms with different
powers of $z$, a set of three coupled
self-consistency equations
\begin{eqnarray}
\alpha_0&=&\frac{-g}{\sqrt{\pi}}\sum_{\zeta}\sum_k
\left[u_{0,k}^\zeta v_{0,k}^{\zeta *}+\frac{1}{2}u_{2,k}^\zeta
  v_{2,k}^{\zeta *}\right.\\
&-&\left.\frac{1}{\sqrt{2}}\left(u_{0,k}^\zeta 
v_{2,k}^{\zeta *}+u_{2,k}^\zeta v_{0,k}^{\zeta *}\right)\right]
\left[1-2n_F(E_{\zeta,k})\right],\nonumber \\
\alpha_2&=&\frac{-g}{\sqrt{\pi}}\sum_{\zeta}\sum_{k}
\left[2u_{1,k}^\zeta v_{1,k}^{\zeta *}-2u_{2,k}^\zeta v_{2,k}^{\zeta *}\right.\\
&+&\left.\sqrt{2}\left(u_{0,k}^\zeta v_{2,k}^{\zeta *}+u_{2,k}^\zeta v_{0,k}^{\zeta *}\right)\right]
\left[1-2n_F(E_{\zeta,k})\right],\nonumber \\
\alpha_4&=&\frac{-g}{\sqrt{\pi}}\sum_{ \zeta}\sum_k
2u_{2,k}^\zeta v_{2,k}^{\zeta *}\left[1-2n_F(E_{\zeta,k})\right].
\end{eqnarray}
We solve this set of 
equations numerically. 

Replacing the sums over $k$ with 
two-dimensional integrals gives rise to equations which are formally
divergent. The ultraviolet divergence has its origin in
approximating the interaction between atoms with a contact
interaction. Most elegantly this divergence is removed by
renormalizing $g$ to two-body scattering matrix, which amounts
to subtracting the divergent part away from the integral.
However, for computational reasons, we remove the divergence 
using a simple Gaussian energy cut-off. The cut-off energy is high
enough so that our results are not
sensitive to the cut-off procedure. 

The integrals are also
infrared divergent and this divergence is due to the
bound state appearing in the two-dimensional problem.
This bound state has an energy 
$\epsilon_0/\hbar\omega_z\sim \exp\left(-\sqrt{2\pi}l_z/|a|\right)$, where $a$ is the three
dimensional scattering length~\cite{Petrov2003a}.
This weakly bound state causes a logarithmic energy dependence for the coupling strength
$g\sim \ln^{-1}\left(\epsilon_0/\epsilon\right)$ which in turn
results in non-separable gap equations. In our case this makes
the computations exceedingly complicated. Furthermore, the bound-state
energy also depends on the axial wave function.
This implies that the relevant bound state energy would
also depend on the harmonic oscillator quantum number $n$ and consequently one
would have to deal with several different coupling strengths. 
However, in the BCS theory, the most interesting effects originate in the
neighborhood of the Fermi surface. System behavior is therefore
largely insensitive to the details of the low energy behavior
of the coupling strength. In fact, we found that 
the use of a sharp low energy cut-off is sufficient for the examples 
presented here. As long as the low energy
cut-off was $\ll 1$, we could change it by an order of magnitude
without affecting the results seriously.
  
\begin{figure}
\includegraphics[width=\columnwidth]{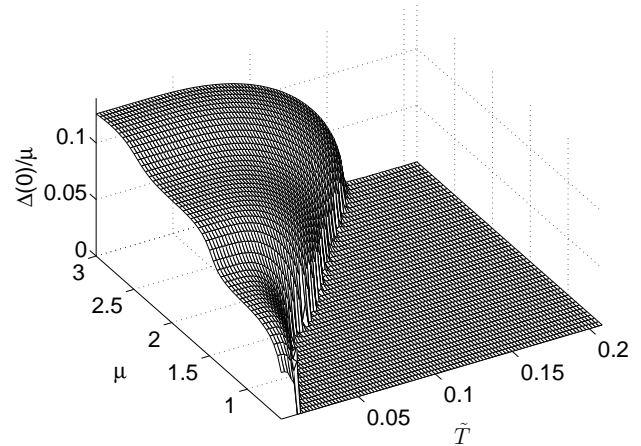}
\caption[Fig1]{Order parameter $\Delta(z)$ at $z=0$ as a function of
dimensionless temperature $\tilde{T}=k_BT/\hbar\omega_z$ 
and chemical potential. We assumed a dimensionless coupling
strength $g=-0.25$ here and in Fig.~2.
}
\label{fig:gapsolution}
\end{figure}

In Fig.~\ref{fig:gapsolution} we show, for $^{6}{\rm Li}$ atoms, an example of the
order parameter in the center of the system 
$\Delta(z=0)$ as a function of temperature and chemical potential.
The gap $\Delta(0)$ increases
with the chemical potential and shows a clear staircase structure
especially around $\mu=3\hbar\omega_z/2$. The behavior around $\mu=5\hbar\omega_z/2$ is 
smoother. Also the critical temperature rises quite suddenly when the
chemical potential is close to the harmonic oscillator levels. This increase 
reflects the abrupt increase of the density of states.
While $\Delta(0)=\alpha_0$ increases monotonically with
chemical potential, this is not generally true for other coefficients
of $\Delta(z)$. 
Having solved the gap equations, the derivative $\partial n_{2D}/\partial\mu$ 
can be easily computed. The staircase
structure one expects for an ideal Fermi gas is still present, but
now the interactions have rounded the steps. 

In a homogeneous superfluid, the energy gap for the
single particle excitations coincides with
the order parameter. In the inhomogeneous case, the order parameter becomes position dependent and 
Andreev bound states (in-gap states)
appear~\cite{deGennes1963a}. 
Note that in the RF-spectroscopy experiments so far, the final 
state was initially empty, i.e.~there is no Pauli blocking for transfer of particles from the lowest momentum states, unlike in superconductor-normal 
metal tunneling experiments. In a strongly interacting Fermi gas, however, the order parameter can be as large as half of the Fermi energy and even 
the lowest momentum states are strongly affected by pairing. 
As a result, it turns out 
that the peak position is of the same order of magnitude as the order parameter~\cite{Chin2004a,Kinnunen2004b}. In contrast, in the BCS limit, pairing 
takes place near the Fermi level and the transfer of particles from the lowest momentum states leads to a very small shift of the
peak position~\cite{Torma2000a}, of the order
$\Delta^2/(2\mu)$. Such small shifts are difficult to observe.
However, in the near future it will be possible to trap stable
mixtures of Fermions in three internal states 
of one atom (or, say, in one state of $^{40}$K and two of $^6$Li). Then, selected transitions can be Pauli blocked by preparing the Fermi level of the 
final state at will. This provides a new degree of control in the spectroscopy of the gas
and the resolution at the BCS limit could be dramatically increased. 
In this way, also the Andreev state energies might be directly observed.
An alternative way to resolve the lowest Andreev state energy could be via suppression of collective mode frequencies~\cite{Ohashi2005a}.

\begin{figure}
\includegraphics[width=0.92\columnwidth]{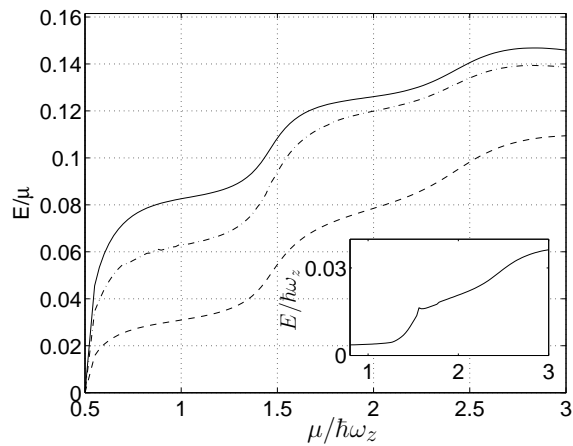}
\caption[Fig2]{The lowest Andreev state single particle energy 
(dashed  line), peak position of the RF-spectrum (dot-dashed line), 
and $\Delta(z=0)$ (solid line) as a function of chemical potential at $T=0$.
The inset shows the peak position when the final state in the RF-spectroscopy is initially empty
while in the main figure equal chemical potentials of the initial and final state were assumed.
}
\label{fig:energygap}
\end{figure}

We calculated the
RF-spectra for our system~\cite{Torma2000a} and from that determined the location of the 
spectral peak as a function of chemical potential.
This peak position and the lowest Andreev state energy together with
the order parameter at $z=0$ are shown
in Fig.~\ref{fig:energygap}. In the main figure, the peak position is calculated by 
assuming Pauli blocking of the final state, i.e.~the final state chemical potential equals the initial state one. 
The spectral peak is closer
to $\Delta(0)$ than to the lowest Andreev state. This trend becomes
more clear as the chemical potential increases, meaning that the experimental signal in RF-spectroscopy from the lowest Andreev state
becomes negligible
when several axial harmonic oscillator states are occupied. This might, however, be changed by the choice of the final state chemical potential. In 
the inset we show the peak position when the final state was initially
empty. The spectral shift is now much smaller.
However, both the main figure and the inset demonstrate that the steps in the order parameter are directly reflected in observable quantities.
This behavior originates from the change in the density of states and demonstrates the many body nature of the pairing.

Probing experimentally the regime we are interested in here requires a
sufficiently tight
axial confinement. In practice, the interesting two-dimensional
density scale is $n_{2D}\sim 1/(\pi l_z^2)$ and when the axial
profile of the cloud is that of the lowest harmonic oscillator state, this
corresponds to
a density scale $n_{3D}\sim 1/(\pi l_z^3)$ at $z=0$. If 
$\omega_z=2\pi\cdot 1000\,{\rm Hz}\times s$, where $s$ is 
dimensionless, we get for $^{6}{\rm Li}$ atoms 
$n_{3D}\sim s^{3/2}\times 1.5\cdot 10^{11}\,1/{\rm cm^3}$.
Assuming that the density in the real experiment is between
$10^{13}\,1/{\rm cm^3}$ and $10^{14}\,1/{\rm cm^3}$, one would
require axial trapping frequencies in the range of 10-100 kilohertz or
more precisely
$s\sim 20\ldots 80$. This is entirely realistic using, for example, 
a one dimensional optical lattice. Due to the larger atomic mass, for Potassium 
the required axial trapping frequency is smaller
by factor of about three. The features in Fig.~\ref{fig:energygap} become of the order 1-10 kHz 
which can be resolved in an experiment.  

In an experiment the cloud is inhomogeneous also
radially. This  inhomogeneouity is expected to 
broaden the staircases further. If the radial density
profile varies on a length scale $R$, we expect a natural
energy scale for inhomogeneity of $\sim \hbar^2/2mR^2$ and 
effects due to it become small if it is much smaller than
the interaction energy scale $\sim |g|\hbar\omega_z=|a|\hbar\omega_z/l_z $. This implies
a condition $(R/l_z)^2\gg (l_z/|a|)$. 
Since the scattering length is tunable using Feshbach resonances this
condition is not difficult to satisfy.
Furthermore, many effects due to the radial density profile can
be included, if necessary, using the local density 
approximation. 

In a purely two-dimensional system one expects a Berezinskii-Kosterlitz-Thouless
transition associated with the proliferation of vortex-antivortex 
pairs. 
This implies non-trivial phase fluctuations of the order parameter $\Delta(z)$
and the absence of long range order (when $T\neq 0$) as required by the
Coleman-Mermin-Wagner-Hohenberg theorem~\cite{ColemanMerminWagnerHohenberg}. Here we have
ignored such fluctuations with the implicit assumption that the modulus
$|\Delta(z)|$ of the order parameter can be reliably calculated 
with the BCS theory. Understanding
how the couplings between harmonic oscillator levels suppress phase
fluctuations and establish true long range order would be an
interesting future extension of the work. 
 
In summary, we have shown how finite size effects appear and can be observed in a superfluid
Fermi gas that is quasi two-dimensional via a tight harmonic confinement in one dimension. We introduced a theoretical approach which 
employs the first few oscillator states. Such an approach is sufficient to describe finite size effects, yet it is simple and transparent, which
should be very useful in studies of phase coherence and fluctuations in dimensional crossovers, of strong interactions, as well as of superfluid
fermion dynamics in this system. 
 
We thank J. Kinnunen for helpful discussions. This work was supported by  
Academy of Finland (project number 106299).


\end{document}